\documentstyle[12pt,aaspp4]{article}

\begin{document}

\title{Broad band X-ray spectrum of Cygnus X-1  }
\author{V.R. Chitnis, A.R. Rao, and P.C. Agrawal}
\affil{Tata Institute of Fundamental Research, Himi Bhabha Road,
Mumbai, 400~005, India}
\affil{e-mail: varsha@tifrvax.tifr.res.in (VRC), arrao@tifrvax.tifr.res.in (ARR)
and pagrawal@tifrvax.tifr.res.in (PCA)}

\begin{abstract}
   We present the hard X-ray (20 $-$ 100 keV) observations of Cygnus X-1
obtained using a large area balloon-borne Xenon filled Multi-anode Proportional
Counter (XMPC) telescope. The observations were carried out during the
$\gamma_2$ state of the source and we obtain a power law photon
index of 1.62$\pm$0.07. To constrain the spectral shape of the
source, we have
analyzed the archival EXOSAT ME argon and GSPC data in the low energies
(2 $-$ 20 keV band)
as well as the archival OSSE data in the high energies (50 $-$ 500 keV).
The data in different energy bands are not obtained in simultaneous
observations, but they pertain to the $\gamma_2$ state of the source.
We have attempted a combined fit to the wide band data using appropriate
mutual detector calibrations. 
This method implicitly assumes that the 
variations in the source intensity in the $\gamma_2$ state is mainly
due to the variations in the normalisations of the spectral components
rather than any change in the 
spectral parameters.
A combined fit to the EXOSAT and XMPC data
(2 $-$ 100 keV) shows that the observed spectrum requires a low
energy absorption corresponding to the Galactic interstellar absorption,
a low energy excess modeled as a blackbody, a narrow emission line due
to iron K$_\alpha$ and a continuum. The continuum can be either modeled as a 
power law
with a reflection bump or a Comptonisation model with an additional bump
which can be modeled as the partial covering with a heavy absorber.
To resolve between these two models, we have attempted a combined fit to the 
2 $-$ 500 keV data obtained from EXOSAT, XMPC and OSSE. We find that a
single Comptonisation model cannot adequately represent the continuum. 
The observed excess is at higher energies (about 100 keV) and it cannot be
modeled as
reflection of power law or Comptonisation model. We find that a two component
Comptonisation model adequately represents the data. We explore the possible
emission region that is responsible for the observed spectrum.


\end{abstract}

\section{\bf Introduction}
 
    Cygnus X-1 is a well-known Galactic black hole candidate
studied extensively by almost every X-ray astronomy experiment
since its discovery. It shows two distinct spectral states, 'low'
and 'high', depending on its soft X-ray flux. Soft X-ray spectrum 
changes drastically between these two states
(with an equivalent power law photon index changing from $\sim$1.5
to $>$ 3-4), whereas hard X-ray (20 $-$ 100 keV)
spectrum is described by a power law with photon index ranging from
1.2 to 2.3
at all times (Liang and Nolan 1984).
    Low state soft X-ray continuum (2 $-$ 10 keV) of Cyg X-1 is 
generally characterized as a power law with a photon index of about 1.5. 
Barr and van der Woerd (1990) 
detected the presence of a soft excess in
a combined fit to  the EXOSAT Transmission Grating Spectrometer (TGS)
data (0.4 $-$ 2 keV) and Gas Scintillation Proportional Counter (GSPC)
data (2 $-$ 12 keV), which they have modeled as a steep power law and a
broad emission feature between 550 eV and 800 eV. Similar results 
were obtained by Balucinska and Hasinger (1991) while analyzing the
EXOSAT Medium Energy (ME) data (1.3 $-$ 15 keV). They modeled the
soft excess as a blackbody with kT $\sim$ 0.3 keV. The BBXRT 
observations probably did not detect this component 
(Marshall et al. 1993). Analyzing the Rosat PSPC data over 0.1 $-$ 2 keV near 
the binary orbital
phase of 0.5, Balucinska-Church et al. (1995) characterized the soft
excess with a blackbody with temperature kT = 0.13$\pm$0.02 keV.

 Barr et al. (1985) discovered iron K$_\alpha$ line in 
Cyg X-1 using the EXOSAT GSPC data. They found it to be red-shifted
(center at 6.2 keV), broad ($\sim$ 1.2 keV) and weak (equivalent
width $\sim$ 120 eV). Kitamoto et al. (1990) detected the iron line
centered at 6.5 keV using the Tenma GSPC data. 
Kitamoto et al. (1984) also detected
an absorption edge at 7.18$\pm$0.18 keV only during a dip, whereas
Balucinska-Church and Barr (1991) saw an edge at 7.1 $-$ 7.8 
keV during persistent emission implying ionisation state between  
neutral to FeXVII. Tanaka (1991) and Ebisawa (1991) modeled
the Cyg X-1 spectrum (from Ginga observations) using reflection of
power-law and Gaussian line and found iron line and edge at 6.4 keV
and 7.2 keV respectively, corresponding to neutral iron. They also
showed that the line is red-shifted (6.1 keV) if the continuum is fitted
by power-law only. Done et al. (1992) analyzed EXOSAT ME Ar,
GSPC and HEAO1-A2 data and fitted it with a Compton reflection of
power-law photons with photon index $\sim$ 1.8, from an ionized 
disk. The observed average iron line energy is 6.44$\pm$0.12 keV and 
the line is found to be narrow.
BBXRT spectrum over 3 $-$ 11.2 keV is consistent with a disk reflection
without indication of ionisation, and with little evidence of line 
emission, with  a  possible exception of a narrow emission line at 6.4 keV
with an equivalent width of 13$\pm$11 eV (Marshall et al. 1993). 
Ebisawa et al. (1996) detected an iron line at 6.4 keV with 
intrinsic width of $\le$ 0.2 keV and equivalent width of 10 $-$ 30 eV
along with a broad edge feature in the ASCA SIS data.

       The hard X-ray spectrum of Cyg X-1 over the energy range of 
20 $-$ 200 keV (Sunyaev and Trumper 1979) was fitted by a thermal
Compton model (Sunyaev and Titarchuk 1980; hereafter referred to as the
CompST model) with an electron temperature of kT$_e$ = 27 keV and 
optical depth $\tau$ = 5, in order to account for the steepening of
the spectrum above 100 keV. While fitting broad band data over the energy
range of 3 keV to 8 MeV, obtained from HEAO $-$ 1, Nolan et al. (1981)
found that a single CompST model is insufficient to fit the data and they
used two Comptonisation models with temperatures kT$_e$ of 15.2 and
41.6 keV, respectively.

       The presence of a 'super-low' state in Cyg X-1, characterized
by a decrease in both the soft and hard X-ray flux, was first detected
by Ling et al. (1983) while analyzing data from JPL high resolution 
gamma ray spectrometer on board HEAO-3. Fitting the data over 48 $-$ 300
keV range they concluded that the hard X-ray spectrum does not change
much between the 'super-low' and the 'low' state, whereas it is quite
different during the 'high' state. In an extension of this work, Ling
et al. (1987) divided the hard X-ray (45 $-$ 140 keV) states of Cyg X-1 
into a $\gamma_1$ state (corresponding to the 'super-low' state), a $\gamma_2$ 
state (the 'normal' state to which the source frequently returns) and 
the infrequently occurring $\gamma_3$ ('flare') state. Salotti et al. (1992)
fitted the 35 $-$ 750 keV data obtained from the SIGMA detectors
on board GRANAT satellite using 
CompST model with kT$_e$ = 62 keV and $\tau$ = 2. Grebenev et al. (1993) 
extended this fit to lower energies in attempting a combined fit for
ART-P (2 $-$ 60 keV) and SIGMA data and found that it overestimated flux
at lower energies, whereas simultaneous fit over the energy range 2 $-$ 600
keV with CompST (kT$_e$ = 37 keV and $\tau$ = 1.5) underestimated the flux at
higher energies ($>$ 100 keV).
%

       D$\ddot{o}$bereiner et al. (1995) fitted HEXE spectrum over 2 $-$ 200 
keV range pertaining to the $\gamma_2$ state of the source with Compton 
reflection of power-law ($\alpha$ = 1.64$\pm$0.01) from cool matter
with covering factor of 0.59$\pm$0.04. This model provided better fit
than a power-law or CompST model to the HEXE data. Phlips et al. (1996) 
fitted Cyg X-1 spectral data over 50 keV $-$ 1 MeV obtained from OSSE
on board CGRO with various models and found that the exponentially truncated 
power-law with a photon index 1.39$\pm$0.02 and cutoff energy kT = 158$\pm$3
keV give a fit to the data better than thermal bremsstrahlung or 
thermal Comptonisation
model. COMPTEL data over the energy range 0.75 $-$ 30 MeV obtained in 
June and August 1991 are described by Wien spectral model (high energy limit
of CompST) with plasma temperatures 192$\pm$27 keV and 204$\pm$21 keV,
respectively (McConnell et al. 1994).

       Most of the attempts so far to explain the energy spectrum of Cyg X-1
were over limited dynamic range and even when data over wider band-widths
were used, they were generally from detectors of poor energy resolution.
To understand the nature of the underlying continuum it is necessary to 
have good energy resolution at lower energies, wider band-width at higher
energies to include the turnover at $>$ 200 keV, and medium energy resolution
at the intermediate (20 $-$ 100 keV) energies. We have achieved this   
by combining data obtained by us in a balloon flight
of Xenon filled Multi-anode Proportional Counters (XMPCs) over energy range 
of 20 $-$ 100 keV, with EXOSAT ME Argon and GSPC data in 2 $-$ 20 keV band
and OSSE data over 50 $-$ 500 keV range. We have carried out a simultaneous fit
to these data sets pertaining to the $\gamma_2$ state of the source.
Since the source generally remains in this state with a flux variation
of less than a factor of 2, we have derived the spectral parameters under
the assumption that the flux variation pertains to only the normalisation
constants of the various spectral components rather than the
spectral  parameters. 

      The paper is organized as follows. In section 2 we describe 
of the XMPC observations, details of the response matrix of the
 detectors and
the results obtained from a spectral fit to the Cyg X-1 data.  
In section 3 we describe a wide band spectral model for Cyg X-1.
The results are discussed in section 4 followed by summary in the
last section.

\section{\bf XMPC instrument details and observations}

     Hard X-ray observations of Cyg X-1 were carried out in a balloon 
flight of a telescope consisting of two xenon filled multi-anode 
proportional counters (XMPCs) each with an area of 1230 cm$^2$, carried out
on 1992 April 5/6 from Hyderabad, India. These detectors have an average
X-ray detection efficiency of about 50\% between 20 and 100 keV. Active
volume of the detectors is divided into three layers, with four anode
cells of cross-sectional area 4.8 cm $\times$ 4.8 cm in each layer. Alternate
anode cells in each layer are joined together. The anode cell assembly is 
surrounded by
veto cells on three sides. In order to reduce background induced by
charged particles, anode cells are operated in mutual anti-coincidence to
reject simultaneous events from different anodes. To avoid rejection of
genuine X-ray events above 34.5 keV, escape gating technique is used, 
which accepts two simultaneous events, provided one of them is in 25 to 
35 keV band, corresponding to the xenon K-shell fluorescent event. In case of 
the escape
gated event, only non-K event is analyzed for the pulse height and the energy
of the K-shell fluorescent event is assumed to be 29.7 keV. 
The veto layer is operated in 
anti-coincidence with the main detection cells to reject background induced by 
charged particles. A mechanical graded slat collimator of tin and copper
restricts the field of view to 5$^\circ$ $\times$ 5$^\circ$ FWHM. For
details of the X-ray telescope refer to Rao et al. (1987, 1991).

     The balloon flight was carried out on 1992 April 5 at 19:01 UT, and the
balloon reached a ceiling altitude corresponding to a residual atmospheric 
column density of
4 gm cm$^{-2}$ at 21:15 UT. The payload is oriented using an alt-azimuth
orientation system with a pointing accuracy of 0.3$^\circ$. Source tracking 
is done according to azimuth and elevation angles stored in an on-board
programmer, which are updated every minute. Cyg X-1 was observed 
continuously for a duration of one hour starting at 1:50 UT on 1992 April
6, followed by 4 cycles of source and background observations, for 20 
and 10 minutes, respectively. Count rate profile of Cyg X-1 obtained from one of
the XMPCs during this balloon flight is shown in Figure 1. The effect of change 
in air mass shows as a decrease in the source count rate away from the 
meridian 
transit time. As can be seen from the figure, the background count rate
remained steady throughout the observations. The gap in the observation
before the meridian transit is the duration for which the source NGC 4151
was being tracked. Aspect calibration by the triangulation method is
attempted several times during the beginning of the observations. From a
detailed fitting of the variation of count rates during aspect calibration
and also with the zenith angle of the source, we estimate that the error
in the orientation of the telescope can contribute to an extra vignetting
correction of about 0.2 (corresponding to an angular offset of 1$^\circ$). 
Increase in the count rate for 
a duration of about 10 minutes at the end of the one hour tracking is due to
a gamma-ray burst. Average Cygnus X-1 count rate near the meridian transit was
25.5 $\pm$ 0.4 counts s$^{-1}$, after subtracting the background count rate.

\subsection{\bf Detector response matrix}

      In order to characterize the energy spectra of X-ray sources it is
necessary to have a detailed knowledge of the response of the detector. 
Figure 2 shows the response of a single anode of  a  detector for fluorescent 
K$_\alpha$ and K$_\beta$ X-rays from Tb.
Various photo-peaks and escape peaks can be seen. The energy resolution of
the detector is $\sim$ 9.5\% at 44 keV and it shows a weak dependence on 
energy. As can be seen from the figure, xenon gas has a very large 
fluorescence yield and to take care of such effects, we have generated response
matrix of the XMPC using a Monte Carlo routine. Inputs to this program
include detector characteristics such as partial pressure of xenon gas in
the detector, layer-wise conversion gain (relation between output pulse 
height and input energy), energy resolution of the detector, 
geometry of the detector, and 
characteristics of event selection logic such as thresholds of K-band,
thresholds of upper and lower level discriminators which defines energy
range of acceptable events. In the simulation program, each photon is tracked
assuming random incidence at the detector surface. Position and layer number
for the interaction is then calculated. Possibility of emission of K-X-ray
is considered and layer number corresponding to interaction of the K-X-rays is
calculated. Taking into account gain and energy resolution of the detector,
the output pulse height is evaluated. Escape gating technique is taken into
consideration by calculating pulse height only for non-K event in case of
simultaneous events. Depending on the type of interaction  and the layer in 
which the 
interaction took place, a layer identification number is generated. 
Simulation is done typically 10000 times for each energy bin of width 1 keV,
between 10 and 130 keV. The corrections due to absorption in the air and the
window material are calculated numerically.

        Various inputs to this routine, e.g. layer-wise gain and energy
resolution of the detector, various event selection logic thresholds etc.,
are computed by calibrating the detectors with various radioactive X-ray
sources of known energies. The observed response of the detector to 
mono-energetic X-rays is compared with the simulated one. It was found that
one of the detectors had better uniformity of gain and overall energy
resolution (10\% FWHM at 60 keV) and the observed spectra from calibration 
sources were found to show good agreement with the predicted spectra correct
to about 2\% in each spectral bin. 

       Figure 3 shows the response of the second layer of the detector A for
Am$^{241}$ radioactive source. Data points with error bars correspond to the observed 
pulse height distribution and the histogram corresponds to the predicted
pulse height distribution obtained by convolving a Gaussian line at 60 keV
with the Monte Carlo simulated response of the detector. Am$^{241}$ photo-peak 
at
60 keV and two escape peaks at 26 and 29.9 keV corresponding to the escape of
xenon K$_\beta$ and K$_\alpha$ X-rays can be seen in the figure.

\subsection{\bf Hard X-ray spectrum of Cyg X-1}

       We have selected data from the continuous tracking of Cyg X-1 near 
meridian transit, from one of the detectors which has better spectral
response, for spectral fitting. 
The spectral files are generated for separate layers and
data for each layer  are re-binned in 42 channels.
       We have used the XSPEC package (Arnaud 1996) for the spectral
fitting. Simultaneous spectral fits for the three layers were carried out.
A power-law with a photon index ($\Gamma$) of 1.62$\pm$0.07 (90\%
confidence limits) gives an acceptable value of $\chi^2_{min}$ of 129 for 125 
degrees of freedom (dof). The observed 20 $-$ 100 keV flux is 1.3 $\times$
10$^{-8}$ ergs cm$^{-2}$ s$^{-1}$, which corresponds to a luminosity of
0.9 $\times$ 10$^{37}$ ergs s$^{-1}$ (for a distance of 2.5 kpc). The
observed flux density at 100 keV is 6.4 $\times$ 10$^{-4}$ photons cm$^{-2}$
s$^{-1}$ keV$^{-1}$. The observed count spectra obtained from layer 1 and
layer 2 and 3 summed together, are shown in fig. 4. The best fit power-law
spectrum, convolved with the detector response is shown as histograms. The
residuals to the fit are shown in lower panel, as contribution to 
$\chi^2$.

      We have also made an attempt to fit other spectral models to the 
XMPC data. Thermal bremsstrahlung model gives a $\chi^2_{min}$ of 126 for 125 dof,
with temperature $>$ 126 keV. The CompST model gives $\chi^2_{min}$ of 123 for
124 dof, with electron temperature kT$_e$ of 25$^{+18}_{-6}$ keV and the optical
depth $\tau$ of 4.8$^{+1.6}_{2.4}$ (90\% confidence errors for 2 free
parameters). Due to limited bandwidth of the data we cannot distinguish 
between any of these models.

      In order to know the exact nature of the continuum, which is not
possible due to the limited dynamic range of the XMPC data, we have extended the
dynamic range by combining XMPC data with EXOSAT ME Argon and GSPC data 
spanning 2 $-$ 20
keV band and OSSE data over 50 $-$ 500 keV. 
The source normally remains in the $\gamma_2$ state, occasionally going to
the super-low ($\gamma_1$) or the flare ($\gamma_3$) state (Ling et al. 
1987).
The flux density  at 100 keV 
derived from
power-law spectral model for XMPC data agrees well with the average value
derived from 10 observations from the SIGMA detector (6.37 $\times$ 10$^{-4}$
photons cm$^{-2}$ s$^{-1}$ keV$^{-1}$) made between 1990 March and 1992 March,
when the source was in $\gamma_2$ state (Laurent et al. 1993), and hence
we can conclude that the source was in the $\gamma_2$ state during our
observations. We have
selected EXOSAT and OSSE data pertaining to $\gamma_2$ state of Cyg X-1 and
carried out simultaneous fit to EXOSAT, XMPC and OSSE data spanning an energy
range of 2 $-$ 500 keV.

\section{\bf Wide band spectrum of Cyg X-1}

\subsection{\bf The EXOSAT data and simultaneous spectral fit to EXOSAT and
XMPC data}

     The EXOSAT archive contains spectral data obtained from medium energy
(ME) argon filled (Ar) detectors and gas scintillation proportional counters
(GSPC). The Ar detectors have an area of $\sim$ 1600 cm$^2$ and they are
sensitive in 1 to 20 keV range (Turner, Smith and Zimmerman 1981). The ME
data is available in 128 pulse height (PH) channels (extending to 60 keV) or
in 64, 32 or 8 channels (extending to 20 keV). The GSPC, although having a
smaller area ($\approx$ 150 cm$^2$), has an energy resolution that is a factor
of 2 better than the Ar detectors (Peacock et al. 1981). The GSPC can operate
at gain 2 covering an energy range of 2 $-$ 16 keV in
256 PH channels or gain 1 covering 4 $-$ 32 keV in 256 PH channels (Peacock 
et al. 1981).

We have used the criteria of Done et al. (1992) for selecting the EXOSAT data,
viz., selecting good quality data without absorption dips, with proper 
background
subtraction and good spectral binning.
There are 5 GSPC and 4 ME spectra which satisfy the criteria. Further, in
order to carry out combined fit for the ME and the GSPC data, we have selected 
simultaneous GSPC and ME observations. This constraint reduces the number of 
data sets
to 3 taken on 1984 July 9 (84/191 $-$ 191$^{th}$ day of 1984), 1984 November
2$-$3 (85/307$-$308), and 1985 September 14 (85/257). These data files are 
referred to here as set-1, set-2 and set-3 respectively, and the number of 
PH channels, corresponding file names in Done et al. are given in Table 1.

We performed a careful check on the energy scale and mutual area calibration
of the two detectors (Ar and GSPC), following the method given in Rajeev et
al. (1994). The Ar detector is the most used and well understood of all the
three non-imaging detectors in the EXOSAT Observatory and we have suitably
adjusted the energy gain of the GSPC detector with respect to the Ar detector.
The area corrections were made on the Ar data and these corrections can be
larger than those found for Cyg X-3 by Rajeev et al. due to differences in
the durations of observations. The relative gain and area corrections are 
also given in Table 1. For set-3 the GSPC gain has to be corrected by 3.2\%.
It was found by Rajeev et al. (1994) that the GSPC gain can vary even within
an observation and since such large corrections are unreliable we have not
used the data from set-3 for further analysis. The set-2 data has better
GSPC gain setting (gain 2) and it has Ar data in finer bins. Further, the hard
X-ray flux density at 100 keV was about 8.4 $\times$ 10$^{-4}$ photons cm$^{-2}$
s$^{-1}$ keV$^{-1}$ in 1984 October (McConnell et al. 1989; Ubertini et al. 
1991). This value is within 25\% of the flux measured by us in 1992 April. 
Hence we can conclude that the set-2 data pertains to the $\gamma_2$ state
of the source and we have used this data for further analysis.

We have attempted a combined fit to the EXOSAT ME Ar data (2 $-$ 20 keV),
GSPC data (5 $-$ 16 keV) and XMPC data (20 $-$ 100 keV).
The ultra-soft component modeled as Gaussian lines in 
Barr and van der Woerd (1990) does not contribute significantly above 2 keV.
We have modeled the low energy component as a blackbody emission. 
The interstellar
neutral absorption in the line of sight  to Cyg X-1 is kept fixed at 7 $\times$
10$^{21}$ cm$^{-2}$ and the absorption cross section given by Morrison and
McCommon (1983) has been used. We have restricted the iron line energy 
between 6.3 and 6.5 keV and line width between 0.1 and 0.2 keV, consistent with
the ASCA measurements (Ebisawa et al. 1996).
For the continuum we have considered the model involving 
the reflection of power-law photons from an ionized disk (pliref) developed
by Done et al. (1992). Since it is known that at higher energies 
Sunyaev-Titarchuk Comptonisation model (CompST) explains the spectral steepening,
we have also considered the CompST model along with  a  variant of CompST :
partial covering (pcfabs) of CompST. 
Using a model consisting of a low energy absorption (abs), blackbody emission
(bbody)  and Gaussian line at soft X-ray energies and pliref, CompST or partial
covering of CompST at higher energies, we have 
attempted a combined fit to EXOSAT Ar, GSPC and XMPC data. For the reflection
model the disk temperature is kept fixed at 10$^5$ K and the disk inclination
is kept fixed at 30$^\circ$ (Done et al. 1992). 

The normalization constant for the XMPC data is kept as a free 
parameter. Best fit parameters of  the  models along with nominal one sigma
errors are given in Table 2. 
The relative normalization constant for XMPC data is determined to be 0.64. 
As can be seen from the table, the CompST model 
alone is inadequate
to describe the continuum (reduced $\chi^2_{min}$ of 1.5). 
The reflection model as well as the partial covering of the CompST give
statistically acceptable fits to the data (reduced $\chi^2_{min}$ close to 1),
though the latter requires an extremely high value of N$_H$ ($>$ 10$^{24}$
cm$^{-2}$) covering 18\% of the flux. Essentially, both the models
imply that over and above a power law continuum there exists an extra emission 
above about 10 keV. 
 
Done et al. (1992) fitted data from EXOSAT ME Ar, GSPC and HEAO-1 A2
over 5 $-$ 50 keV with a model consisting of low energy absorption, iron
K$_{\alpha}$ line modeled with Gaussian and reflection of power-law photons
from ionized accretion disk (pliref). They have fitted data from 
various detectors separately, whereas we have carried out simultaneous   
spectral fit to EXOSAT ME Ar, GSPC and XMPC data spanning much broader 
energy range of 2 $-$ 100 keV. Model parameters obtained here are consistent
with those obtained by Done et al. 

In Figure 5 the observed count rate spectrum for Ar, GSPC
and XMPC (layer 1 and layer 2+3) are shown. The best fit reflection model,
convolved with the individual responses is shown as histogram. The residuals 
to the data are shown in the lower panel of the figure, as contribution to
$\chi^2$. The deconvolved spectra are shown in Figure 6. The blackbody 
component, the Gaussian line  and the reflection component are shown separately 
as histograms. For clarity, the unfolded data points are shown in a non-overlapping
manner.

 
\subsection{\bf The OSSE data and simultaneous spectral fit to XMPC and OSSE
data}

The spectrum of Cyg X-1  below 100 keV is essentially a power-law (see Figure
6). The spectral turnover occurs above about 200 keV and it is essential
to include high energy data for a proper modeling of the continuum.
For this purpose we have analysed the archival data obtained from
the Oriented Scintillation Spectrometer Experiment (OSSE) on-board Compton
Gamma Ray Observatory (CGRO). The OSSE consists of four identical NaI-CsI 
phoswich detectors sensitive to gamma rays with energy 50 keV $-$ 10 MeV. 
Each detector
has an effective area of $\sim$ 500 cm$^2$ at 511 keV (Johnson et al. 1993).
 Cyg X-1 was observed by
OSSE on 17 occasions from 1991 to 1995. We have extracted the longest data 
stretch, with 100 keV flux density within 10\% of XMPC from OSSE archives. These
observations were carried out from  1991 May 30 to 1991 June 7
 (1991/150$-$158), in viewing period
2. Total duration of observations is 24.5 $\times$ 10$^4$ seconds and 
the observed flux density at
100 keV is 6.04 $\times$ 10$^{-4}$ cm$^{-2}$ s$^{-1}$ keV$^{-1}$. 
The 45 $-$ 140 keV flux obtained from an analysis of the OSSE data shows that
the source remained in between the historic $\gamma_1$ and $\gamma_2$ state,
except for a deep super-low state in the beginning of 1994 (see Fig 1. of
Phlips et al. 1996). Since this conclusion is in contradiction with the 
conclusion obtained by Ling et al. that the source generally remains
in $\gamma_2$ state, we have obtained the archival BATSE earth occultation
data. It is found that between 1991 and 1993, the source had an average 
45 $-$ 140 keV flux of 0.11, varying between 0.07 and 0.18 with a standard
deviation of 0.02, without showing any indication of a trend in the light 
curve (fluxes are in units of photons cm$^{-2}$ s$^{-1}$). Since this behavior
is similar to  the $\gamma_2$ behavior, we conclude that the source was in 
the $\gamma_2$ state during 1991-1993. The differences in the absolute values
of the fluxes could be due to the different overall calibration of the 
instruments (the average BATSE flux is 25\% lower than the average $\gamma_2$
flux obtained by HEAO-3).

We have carried out a simultaneous spectral fit to the XMPC and OSSE data, with
the relative normalization as a free parameter, with various spectral models.
We have selected OSSE data from 50 to 500 keV. We have 
tried to fit this data over energy range 20 $-$ 500 keV with Sunyaev-Titarchuk
Comptonisation (CompST) model and power law with exponential
cutoff. The best fit model parameters along with the
nominal one sigma  errors are given in 
Table 3.
It can be seen from the table that a single CompST is totally inadequate
to explain the 20 to 500 keV spectrum of Cyg X-1 
(reduced $\chi^2_{min}$ $>$ 10).
The cutoff power law model too is statistically unacceptable to fit the data
(reduced $\chi^2_{min}$ $>$ about 2). On the other hand, the 
continuum over 20 $-$ 500 keV band is best explained by two CompSTs, with 
temperatures (kT$_e$) 77.5 and 28.4 keV and optical depths ($\tau$) 2.0 and 11.1, 
respectively. Relative normalization  of the OSSE detector with respect to 
XMPC is determined to be 0.85. 
In Table 3, as well as in the subsequent tables, errors in the parameter 
values are not given whenever reduced $\chi^2_{min}$ is greater than 2.

\subsection{\bf Wide band (2 $-$ 500 keV) spectrum}

We have attempted a wide band spectral fit over 2 $-$ 500 keV, combining
EXOSAT Ar, GSPC, XMPC and OSSE data, using the above mentioned spectral 
models. 
The low energy components are modeled as Galactic absorption,
blackbody and Gaussian line. A single CompST for the high energy
continuum is inadequate to fit the data (reduced $\chi^2_{min} \sim$ 6.5).
Reflection of a power-law is also inadequate since there is a spectral
turnover above 200 keV. We have also attempted to fit a model consisting
of reflection of a CompST spectrum, which,  we find, over-predicts the
spectrum above 200 keV. A model consisting of two CompSTs, however, gives
a reasonable value for the reduced $\chi^2_{min}$ (1.23).  
It is known that Cyg X-1
low energy spectrum has an edge like feature near 7.5 keV (Ebisawa et al. 
1996), which was taken into account by the reflection spectrum in our
combined fitting of the EXOSAT and XMPC data (see Figure 6). To explicitly
account for this discontinuity, we also included an edge feature in the
model. The low energy component (which is modeled as a blackbody) generally
shows a tendency to be correlated with the low energy absorption. To account
for this, we also let the absorption vary. These modifications in the model
improved the $\chi^2_{min}$ by 64 (reduced $\chi^2_{min}$ 1.09). 
The best fit spectral model consists of the soft excess modeled
as a blackbody emission with temperature (kT$_{bb}$) 0.33$\pm$0.04 keV, iron line
modeled as a Gaussian with energy 6.47$\pm$0.15 keV and width  0.1
keV, continuum consisting of two Sunyaev-Titarchuk inverse Compton models
with electron temperatures kT$_e$ 80.2$\pm$1.7 and 29.7$\pm$0.6, with
optical depths 1.96$\pm$0.04 and 8.3$\pm$0.8, respectively, along 
with line of sight absorption of (2.34$\pm$0.31) $\times$
10$^{22}$  cm$^{-2}$. The derived edge energy is 7.76$\pm$0.12 keV and
the equivalent width of the iron line is 17.2 eV.
The best fit spectral parameters for this model 
and also for a model where continuum is modeled by single CompST
are given in Table 4. The normalisation constant for the
blackbody radiation is the radius of the emitting region  
 and for the best-fit model it is 75 km 
for a distance of 2.5 kpc and the blackbody luminosity is 8 
$\times$ 10$^{36}$ ergs s$^{-1}$
(or 13 $\times$ 10$^{37}$ ergs s$^{-1}$ for a distance of 10 kpc).
Comparison
of these parameters with those obtained for combined fit to XMPC and OSSE 
data (Table 3) shows that the parameters are consistent in both cases and are
better constrained in 2 $-$ 500 keV spectral fit, because of the larger dynamic
range. The observed count rate spectrum along with the best fit model
spectrum convolved with individual detector response matrices are shown in
Figure 7. Residuals are shown in the lower panel of the figure as
contribution to the $\chi^2$.    The deconvolved spectrum is shown in Figure 8
along with the contribution from the individual model components. For the
sake of clarity only EXOSAT Me Ar, XMPC top layer and OSSE data are
shown here in a non-overlapping manner.
 
\section{\bf Discussion}

Recently Gierlinski et al. (1997) have made a simultaneous spectral fit 
to the 3 $-$ 500 keV data (with a break between 30 and 50 keV) obtained
from the Ginga and OSSE observations. 
The complete spectrum of Cyg X-1 presented here, however, is the first attempt
to fit the spectrum in a very wide band (2 $-$ 500 keV) without any break
in the data. The methodology adopted here implicitly  makes the assumption that
the factor of 2 variability observed in the $\gamma_2$ state is due to
changes in the normalisation constants of the spectral 
components rather than any change in the spectral parameters themselves. 
This assumption is vindicated from the
fact that acceptable values of $\chi^2_{min}$ are obtained for the complete spectral
fit. Further, the data used in the present work has been utilised to examine 
the transition disk model (Misra et al. 1997a) and the conclusions did
not change when simultaneous data was used to examine the same model
(Misra et al. 1997b). The relative normalisation for the different detectors
are used as parameters to be fitted in the fitting procedure. The differences
in normalisation (36\% between EXOSAT and XMPC and 15\% between OSSE and XMPC)
are of the same order as the variations seen in the $\gamma_2$ state. Further,
when a $\chi^2$ contour plot for the normalisation of XMPC with respect to the
OSSE data  and other 
parameters like $\tau$ is made, it is found that the normalisation (85\%) is 
constrained within $\pm$1\%.

When the high energy continuum is adequately modeled,
the parameters obtained for the low energy are consistent with those 
found while analyzing higher resolution data like ASCA and ROSAT. 
The equivalent width of the iron line obtained from the ASCA data
(13 $-$ 40 eV) compares well with the result presented here (17 eV).
The blackbody luminosity derived here 
(13 $\times$ 10$^{37}$ ergs s$^{-1}$ for a distance of 10 kpc) 
is comparable to the value 
of 7.5  $\times$ 10$^{37}$ ergs s$^{-1}$
derived from the ROSAT data by Balucinska-Church et al. (1995).
The interstellar absorption derived here is a factor of 4-5 higher compared
to that obtained from the ROSAT data (which is the same as the  Galactic 
absorption). As pointed out earlier, for the energy band under consideration
($>2$ keV) interstellar absorption and blackbody temperature are related to each other 
and we can only conclude that there exists a soft excess the magnitude of
which is comparable to that obtained from earlier studies.

The high energy continuum is adequately modeled with the final reduced 
$\chi^2_{min}$ of 1.09. 
Since there is a clear break
in the spectral shape near 300$-$500 keV, it is evident that power law or
reflection of power law cannot adequately explain the continuum. Further,
a single component CompST is also not adequate to fit the continuum mainly
because of an excess emission near about 100 keV. We find that this excess cannot
be modeled as a reflection of CompST model: to provide  high reflection at 
100 keV, it requires that the reflection component has to be dominant all 
the way up to 300 keV, which is not seen in the data (see Figure 8).

 Cutoff power-law produces satisfactory fit to the data  above 50 keV (better 
than a single temperature CompST). It, however, cannot explain the continuum
below 20 keV. Best fit cutoff-power-law in hard X-ray band underestimates
the flux when extrapolated below 20 keV. Soft X-ray band requires 
steep power law (index $\sim$ 1.7) and hence a cutoff power-law cannot
explain the continuum completely.
 Phlips et al. (1996) have modeled OSSE data with an exponentially truncated
power-law with reflection. This model, however, does not
explain the continuum at lower energies, since power-law with photon index 
of 0.95 is not
consistent with the low energy data. 

There were a few attempts to model the  wide band X-ray  spectrum of Cyg X-1.
Haardt et al. (1993) showed that a simulated spectrum of reflection of high 
temperature CompST (kT$_e$ of 153 keV) agrees reasonably well with a low 
temperature CompST (kT$_e$ of 63 keV) above 20 keV. What we find here is that
when complete data is taken from 2 keV to 500 keV, single CompST leaves out
a residual whose spectral shape is unlike a reflection component, but agrees
reasonably well with another CompST component. Grebenev et al. (1993) on the
other hand considered 3 $-$ 1300 keV data from the ART-P and SIGMA on-board the
GRANAT satellite. They found that single CompST does not give an acceptable
fit to the data. They, however, could reproduce the overall shape of the
spectrum using a Monte-Carlo simulated Comptonisation spectra from an accretion
disk, though the value of $\chi^2_{min}$ is not mentioned. 

The results obtained by Gierlinski et al. (1997) using the simultaneous
Ginga and OSSE data agrees with the broad conclusions obtained here viz.,
the wide band spectra demands an extra component near 100 keV. This component
can be fit with several different analytical models. What the present work
specifically
showed is that the reasonable energy resolution  in the medium energy band
(20 $-$ 100 keV) is important to constrain this component and the 
good energy resolution at low energies is also important to analytically
formulate this component since the extra spectral component has sufficient
flux below 10 keV. 
A simultaneous observation with
ASCA and XTE and a rigorous spectral modeling will help in clarifying
the spectral shape of Cyg X-1.

As pointed out by Grebenev et al. (1993), it may be a little naive to expect
a single temperature plasma in an accretion disk. Further, at energies near
511 keV one has to include the relativistic Klein-Nishima cross-section of
Compton scattering, which may change the spectral shape at these energies.
The spectral shape obtained
here, however, gives an extremely satisfactory fit to the data all the way
from 2 keV to 500 keV.  In the following we explore the various spectral models
obtained here in light of the accretion disk theory around black holes
developed by Chakrabarti and Titarchuk (1995).

Chakrabarti and Titarchuk (1995) have taken a complete solution of viscous
transonic equations and demonstrated that the accretion disk has a highly
viscous Keplerian part which resides on the equatorial plane and a sub-Keplerian 
component which resides above and below it. The sub-Keplerian component can
form a standing shock wave (or, more generally, a centrifugal barrier
supported dense region)
 which heats up the disk to a high 
temperature. The X-ray spectrum emitted from disks has various components:
the Shakura-Sunyaev disk emission is at low energies and it may be identified
with the low energy blackbody component. The bulk of the
emission comes from cooling emission from Comptonisation which may be
approximated to the CompST model derived here with a temperature of
80 keV. The second CompST model may be an approximation for the hard radiation
reflected from the Shakura-Sunyaev disk along the observer. Another
similarity between the the fitted model and the Chakrabarti-Titarchuk model
is the very low equivalent width for the iron line.
Hence it appears that the various spectral components obtained in the
present study can be identified with the spectral models derived by
Chakrabarti and Titarchuk (1995). It will be interesting to directly fit 
the Chakrabarti and Titarchuk (1995) model to the data.





\section{\bf Conclusions}

We have analysed a wide band X-ray spectrum of Cyg X-1 and have obtained 
a statistically acceptable fit to the 2 $-$ 500 keV data. For this purpose,
we have made hard X-ray observations using a balloon-borne large area xenon
filled multi-anode proportional counter (XMPC) telescope. The response matrix
of the detectors were calculated using a Monte-Carlo routine and the
systematic errors in the data are brought down to a very low level. To extend
the bandwidth, archival EXOSAT and OSSE data  were obtained and a combined
spectral fit to the data was attempted.  
The data in different energy bands, though  not simultaneous, 
pertain to the $\gamma_2$ state of the source.
We have assumed that the 
variations in the source intensity in the $\gamma_2$ state is mainly
due to the variations in the normalisations of the spectral components
rather than any change in the 
spectral parameters.
The main conclusions of this work are:

1. Statistically acceptable spectral fits were obtained using a model 
consisting of a i) interstellar absorption and low energy excess,
ii) iron line and iron absorption edge, iii) continuum extending 
above 500 keV, and iv) an excess mainly near the energy region of 100 keV.
The value of reduced $\chi^2_{min}$ is 1.09 for 416 degrees of freedom.

2. The low energy excess can be modeled as a blackbody with temperature
0.33 keV and luminosity of 8 $\times$ 10$^{36}$ ergs s$^{-1}$. 
The interstellar
equivalent neutral hydrogen column density is 2.3$\pm$0.3 $\times$ 10$^{22}$
cm$^{-2}$.

3. The derived values of line parameters are : the line energy 6.47$\pm$0.15 
keV,
 the
equivalent width of the line  17 eV and edge energy 7.76$\pm$0.12
keV.

4. The continuum can be explained by the Comptonisation model (Sunyaev 
and Titarchuk 1980) with an electron temperature of 80.2$\pm$1.7 keV
and optical depth of 1.96$\pm$0.04.

5. The excess at higher energies can be explained as another CompST model
with an electron temperature of 29.7$\pm$0.6 keV
and optical depth of 8.3$\pm$0.8.

6. All these components have strong similarities with the spectral components
predicted using the black-hole accretion model developed 
by Chakrabarti and Titarchuk (1985).

\begin{acknowledgements}
It is a pleasure to acknowledge the contribution of Shri M.R. Shah,
Electronics Engineer-in-charge of this experiment, in the design, 
development and testing of the XMPC payload. We are also thankful to Shri 
D.K. Dedhia, Shri K. Mukherjee, Shri V.M. Gujar and Shri S.S. Mohite for
their support in the fabrication of the payload. We thank the Balloon
Support Instrumentation Group and the Balloon Flight Group for 
providing telemetry and telecommand packages and successfully conducting the
balloon flight. Thanks are due to the EXOSAT database team for the archive
and the CGRO team for the OSSE  and the BATSE archives.
\end{acknowledgements}

{\bf References}

\begin{itemize}
\item[] Arnaud, K. A.,  1996, in: Jacoby G.H., Barnes J. (eds.) Astronomical
Data Analysis Software and Systems V. ASP Conf. Series Vol. 101, San
Francisco, p17.
\item[] Balucinska-Church, M. Belloni, T., Church, M.J. and 
Hasinger, G. 1995, A\&A, 302, L5.
\item[] Balucinska-Church, M. and Barr, P. 1991,
 In {\it Iron Line 
Diagnostics in X-ray Sources}, Ed. A. Treves, G.C. Perola and L. Stella,
Springer-Verlag, 130p.
\item[] Balucinska M. and Hasinger G. 1991, A\&A, 241, 439. 
\item[] Barr, P., White, N.E. and Page, C.G.  1985, MNRAS, 216, 65p.
\item[] Barr, P. and van der Woerd H. 1990, ApJL, 352, L41.
\item[] Chakrabarti, S.K., and Titarchuk, L.G., 1995, ApJ, 455, 623. 
\item[] D$\ddot{o}$bereiner, S., Englhauser, J., Pietsch, W., Reppin, C., Trumper, J.,
Kendziorra, E., Kretschmar, P., Kunz, M., Maisack, M., Staubert, R.,
Efremov, V. and Sunyaev, R.  1995, A\&A, 302, 115.
\item[] Done, C., Mulchaey, J.S., Mushotzky, R.F. and Arnaud, K.A.
1992, ApJ, 395, 275. 
\item[] Ebisawa, K.  1991, PhD Thesis, University of Tokyo, Japan.
\item[] Ebisawa, K., Ueda, Y., Inoue, H., Tanaka, Y. and White, N.E.  1996,
ApJ, 467, 419.
\item[] Gierlinski, M., Zdziarski, A. A., Done, C., Johnson, W. N.,
Ebisawa, K., Ueda, Y., Haardt, F., Phlips, B. 1997, MNRAS, 288, 958.
\item[] Grebenev, S., Sunyaev, R.A., Pavlinsky, M., Churazov, E., Gilfanov, M.,
Dyachkov, A., Khavenson, N., Sukhanov, K., Laurent, P., Ballet, J.,
Claret, A., Cordier, B., Jourdain, E., Niel, M., Pelaez, F. and
Schmitz-Fraysse, M.C. 1993, A\&ASS, 97, 281. 
\item[] Haardt F., Done, C., Matt, G. and Fabian, A.C. 1993, ApJL, 411, L95.
\item[] Johnson, W.N., Kinzer, R.L., Kurfess, J.D., Strickman, M.S.,
Purcell, W.R., Grabelsky, D.A., Ulmer, M.P., Hillis, D.A.,
Jung, G.V., Cameron, R.A. 1993, ApJS, 86, 693.
\item[] Kitamoto, S., Miyamoto, S., Tanaka, Y., Ohashi, T., Kondo, Y., 
Tawara, Y. and Nakagawa, M.  1984, PASJ, 36, 731.
\item[] Kitamoto, S., Takahashi, K., Yamahita, K., Tanaka, Y. and Nagase, F.
1990, PASJ, 42, 85.
\item[] Laurent, P., Claret, A., Lebrun, F., Paul, J., Dennis, M., Barret, D.,
Bouchet, L., Mandrou, P., Sunyaev, R.A., Curazov, E., Gilfanov, M., 
Khavenson, N., Dyachkov, A., Novikov, B., Kremnev, R. and Kovtunenko, V.
1993, Adv. Space Res., 13, (12), 139.
\item[] Liang, E.P. and Nolan P.L. 1984, Space Sci. Rev., 38, 353.
\item[] Ling , J.C., Mahoney, W.A., Wheaton, W.A. and Jacobson, A.S. 
1983, ApJ, 275, 307.
\item[] Ling , J.C., Mahoney, W.A., Wheaton, W.A. and Jacobson, A.S. 
1987, ApJL, 321, L117.
\item[] Marshall, F.E., Mushotzky, R.F., Petre, R. and 
Serlemitsos, P.S.  1993, ApJ, 419, 301.
\item[] McConnell, M.L., Forrest, D.J., Owens, A., Dumphy, P.P.,
Vestrand, W.T., Chupp, E.L. 1989, ApJ, 343, 317.
\item[] McConnell, M., Forrest, D., Ryan, J., Collmar, W., Schonfelder, V.,
Steinle, H., Strong, A., van Dijk, R., Hermsen, W., and Bennett, K.,
1994, ApJ, 424, 933.
\item[] Misra, R., Chitnis, V.R., Melia, F., Rao, A.R. 1997a, ApJ, in press.
\item[] Misra, R., Chitnis, V.R., Melia, F.  1997b, ApJ, submitted.
\item[] Morrison, R. and McCammon, D. 1983, ApJ, 270, 119.
\item[] Nolan, P.L., Gruber, D.E., Knight, F.R., Matteson, J.L., Rothschild,
R.E., Marshall, F.E., Levine, A.M. and Primini, F.A. 1981, Nat, 293, 275.
\item[] Peacock, A., Andresen, R.D., Manzo, G., Taylor, B.G., 
Villa, G., Re, S., Ives, J.C., Kellock, S. 1981, Space Sci. Rev., 30, 525.
\item[] Phlips B.F., Jung, G.V., Leising, M.D., Grove, J.E., Johnson, W.N.,
Kinzer, R.L., Kroeger, R.A., Kurfess, J.D., Strickman, M.S., Grabelsky, D.A.,
Matz, S.M., Purcell, W.R., Ulmer, M.P., and McNaron-Brown, K. 1996, ApJ, 465, 907.
\item[] Rajeev, M.R., Chitnis, V.R., Rao, A.R. and Singh, K.P. 1994, ApJ, 424, 376.
\item[] Rao, A.R., Agrawal, P.C., Manchanda, R.K. and Shah, M.R. 1987,
Adv. Space Res. 7, (7), 129.
\item[] Rao, A.R., Agrawal, P.C. and Manchanda, R.K. 1991,
 A\&A, 241, 127.
\item[] Salotti, L., Ballet, J., Cordier, B. et al. 1992, A\&A, 253, 145.
\item[] Sunyaev, R.A. and Trumper, J. 1979, Nat, 279, 506.
\item[] Sunyaev, R.A. and Titarchuk L.G. 1980, A\&A, 86, 121.
\item[] Tanaka, Y.  1991,
 In {\it Iron Line 
Diagnostics in X-ray Sources}, Ed. A. Treves, G.C. Perola and L. Stella,
Springer-Verlag,  98p.
\item[] Turner, M.J.L., Smith, A. and Zimmermann, H.U. 1981, Space Sci. Rev.,
30, 513.
\item[] Ubertini, P., Bazzano, A., Perotti, F., Quadrini, E., Court, A.,
Dean, A.J., Dipper, N., Lewis, R., Bassani, L. and Stephen, J.B. 1991,
ApJ, 366, 544.
\end{itemize}

\begin{table}
\begin{center}
\caption{Details of the EXOSAT observations of Cyg X-1} 
\begin{tabular}{lllllllll}
&~~&&&&&&&\\
\hline
&~~&&&&&&&\\
& Observation && File name$^1$ & Duration & PH & Area & Gain & Gain \\
& date && & (s) & channels & correction & & correction \\
&~~&&&&&&&\\
\hline
&~~&&&&&&&\\
set-1 & 1984 Jul 9 & Ar  & 02me  & 1357  & 32  & -2.0\% & - & - \\
      & (84/191)   & GSPC & 08   & 19790  & 256 & - & 1 & 0.0\% \\
&~~&&&&&&&\\
set-2 & 1984 Nov 2 & Ar  & 17me  & 4500  & 64  & -4.0\% & - &  - \\
      & (84/307)   & GSPC & 13   & 5568  & 256  & - & 2 & 0.0\% \\
&~~&&&&&&&\\
set-3 & 1985 Sep 14 & Ar & 15me & 10897.5 & 64 & 7.0\%  & - & - \\
      & (85/257)    & GSPC& 09  & 27969   & 256 & - & 1 & -3.2\% \\
&~~&&&&&&&\\
\hline
&~~&&&&&&&\\
\end{tabular}

$^1$File name, as given in Done et al. (1992)
\end{center}
\end{table}

\begin{table}
\begin{center}
\caption{Spectral parameters of Cyg X-1 in 2 $-$ 100 keV band} 
\begin{tabular}{lllll}
&~~&&&\\
\hline
&~~&&&\\
Model & Parameters$^1$ & value & value & value\\
component & & & &\\
&~~&&&\\
\hline
&~~&&&\\
abs & N$_H$ (10$^{22}$ cm$^{-2}$) & 0.7 & 0.7 & 0.7 \\
&~~&&&\\
bbody & kT$_{bb}$ (keV)& 0.14$\pm$0.03  & 0.14$\pm$0.03  & 0.14$\pm$0.03 \\
& R (km) & 1564$\pm$220  & 2056$\pm$290   & 2500$\pm$350  \\
&~~&&&\\
Gaussian & E$_{Line}$ (keV) & 6.40$\pm$0.15& 6.42$\pm$0.07 & 6.52$\pm$0.06\\
& $\sigma$ (keV) & 0.1$\pm$0.9 & 0.2$\pm$0.3 & 0.2$\pm$0.2\\
& K$_{Line}$ & 1.73$\pm$0.99  & 4.00$\pm$0.71 & 6.44$\pm$0.93 \\
&~~&&&\\
pliref & $\alpha$ & 1.84$\pm$0.01 &&\\
& $\Omega/2\pi$ & 0.73$\pm$0.07  &&\\
& $\xi$ & 56.0$\pm$26.6 && \\
& norm & 2.70$\pm$0.04 && \\
&~~&&&\\
CompST & kT$_e$ (keV)& & 30.9$\pm$4.3 & 41.9$\pm$17.3 \\
       & $\tau$ & & 3.8$\pm$0.3 & 3.0$\pm$0.7 \\
       & norm   & & 2.40$\pm$0.02  & 3.20$\pm$0.08 \\
&~~&&&\\
pcfabs & N$_H$ (10$^{22}$ cm$^{-2}$)   & & & 262$\pm$25 \\
       & cov. fract. & & & 0.180$\pm$0.014 \\
&~~&&&\\
\hline
&~~&&&\\
$\chi^2_{min}$ (dof) & & 369.4 (341) & 509.1 (342) & 369 (340)\\ 
&~~&&&\\
\hline
\end{tabular}
$~~$

$^1$The normalisation constants for the various components are : 
radius of emitting region of the blackbody for a distance of 2.5 kpc; in
units of 10$^{-3}$ 
photons cm$^{-2}$  s$^{-1}$ 
(total) for the Gaussian line;
and  photons cm$^{-2}$  s$^{-1}$  for CompST and pliref.
\end{center}
\end{table}

\begin{table}
\begin{center}
\caption{Spectral parameters of Cyg X-1 in 20 $-$ 500 keV band}
\begin{tabular}{lllll}
&~~&&&\\
\hline
&~~&&&\\
Model  &  Parameters$^1$ & value & value & value \\
component &&&&\\
&~~&&&\\
\hline
&~~&&&\\
CompST & kT$_e$ (keV) & 52.5  & 77.51$\pm$1.79 & \\
       & $\tau$       & 2.9 & 2.01$\pm$0.04 & \\
       & norm         & 2.0 & 2.71$\pm$0.30 & \\
&~~&&&\\
CompST & kT$_e$ (keV) & &  28.45$\pm$1.26 & \\
       & $\tau$       & &  11.1$\pm$5.5 & \\
       & norm         & &  (1.52 $\pm$ 2.21) $\times$ 10$^{-2}$ & \\
&~~&&&\\
cutoffpl & $\alpha$    & & & 0.95 \\
        & cutoff energy (keV) & & & 123.1 \\
        & norm        & & & 0.21 \\
&~~&&&\\
\hline
&~~&&&\\
$\chi^2_{min}$ (dof) &      & 2095 (204) & 209.4 (201) & 458.4 (204) \\
&~~&&&\\
\hline
\end{tabular}
$~~$

$^1$The normalisation constants for the various components are  
in units of photons cm$^{-2}$  s$^{-1}$. 
\end{center}
\end{table}

\begin{table}
\begin{center}
\caption{Spectral parameters for wide band spectrum of Cyg X-1 (2 $-$ 500 keV)}
\begin{tabular}{lllll}
&~~&&&\\
\hline
&~~&&&\\
Model & Parameters$^1$ & value & value & value \\
component & & & & \\
&~~&&&\\
\hline
&~~&&&\\
abs & N$_H$ (10$^{22}$ cm$^{-2}$) &  0.7 & 0.7 & 2.34$\pm$0.31\\
&~~&&&\\
bbody & kT$_{bb}$ (keV) & 0.13 & 0.23$\pm$0.03  & 0.33$\pm$0.04 \\
      & R (km) & 4330 & 120$\pm$13 & 75$\pm$8 \\
&~~&&&\\
Gaussian & E$_{Line}$ (keV) & 6.33 & 6.477$\pm$0.063 & 6.47$\pm$0.15 \\
         & $\sigma$ (keV)   & 0.2  & 0.2 & 0.1$\pm$0.5\\
         & K$_{Line}$ &  5.23  & 4.17$\pm$0.82
          & 1.6$\pm$1.1 \\
&~~&&&\\
Edge & E (keV) & & & 7.76$\pm$0.12 \\
     & $\tau_{max}$ &&& (6.1$\pm$0.9) $\times$ 10$^{-2}$ \\
&~~&&&\\
CompST & kT$_e$ (keV) & 53.6 & 79.55$\pm$1.59 & 80.2$\pm$1.7  \\
       & $\tau$       & 2.8 & 2.00$\pm$0.03 & 1.96$\pm$0.04\\
       & norm         & 2.2 & 2.50$\pm$0.02 & 2.72$\pm$0.06 \\ 
&~~&&&\\
CompST & kT$_e$ (keV) & & 29.36$\pm$0.53 & 29.7$\pm$0.6 \\
       & $\tau$       & & 8.12$\pm$0.56 & 8.3$\pm$0.8 \\
       & norm         & & (4.4$\pm$1.0) $\times$ 10$^{-2}$
  &  (3.9$\pm$1.2) $\times$ 10$^{-2}$\\
&~~&&&\\
\hline
&~~&&&\\
$\chi^2_{min}$ (dof) & & 2727 (422) & 516 (419) & 452 (416)\\
&~~&&&\\
\hline
\end{tabular}
$~~$

$^1$The normalisation constants for the various components are : 
radius of emitting region of the blackbody for a distance of 2.5 kpc; in
units of 10$^{-3}$ 
photons cm$^{-2}$  s$^{-1}$ 
(total) for the Gaussian line;
and  photons cm$^{-2}$  s$^{-1}$  for CompST and pliref.
\end{center}
\end{table}

\clearpage

\begin{figure}
\vskip 5.5cm
\includegraphics{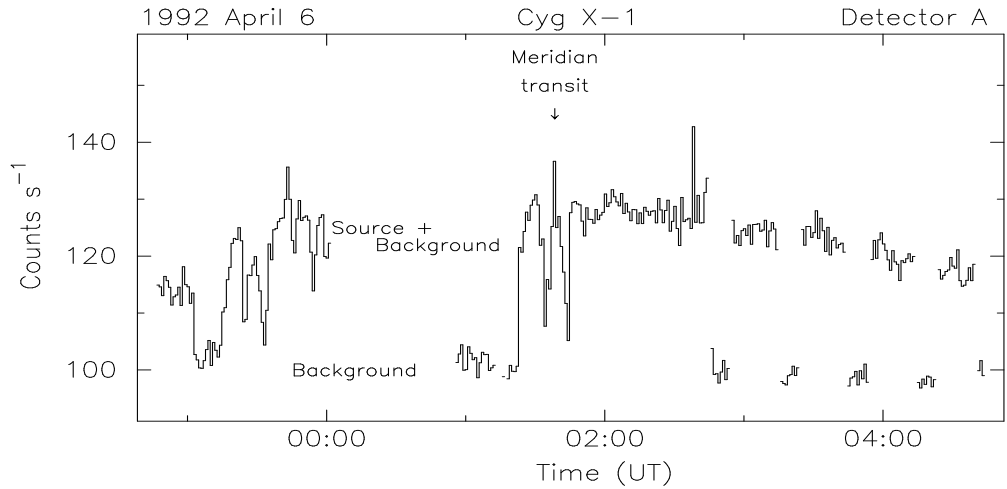}
Š\caption[]
{ The count rate profile of Cyg X-1 obtained from a 
balloon flight carried out on 1992 April 5/6 using the XMPC detector. 
The source and background observations are marked in the figure.
The decrease in the source count rate after the meridian transit is due
to increase in the air mass. During the initial part of the observation
the source was scanned across for aspect calibration. The increase in the
count rate around 02:50 is due to a gamma-ray burst. }
\end{figure}

\begin{figure}
\vskip 7cm
\includegraphics{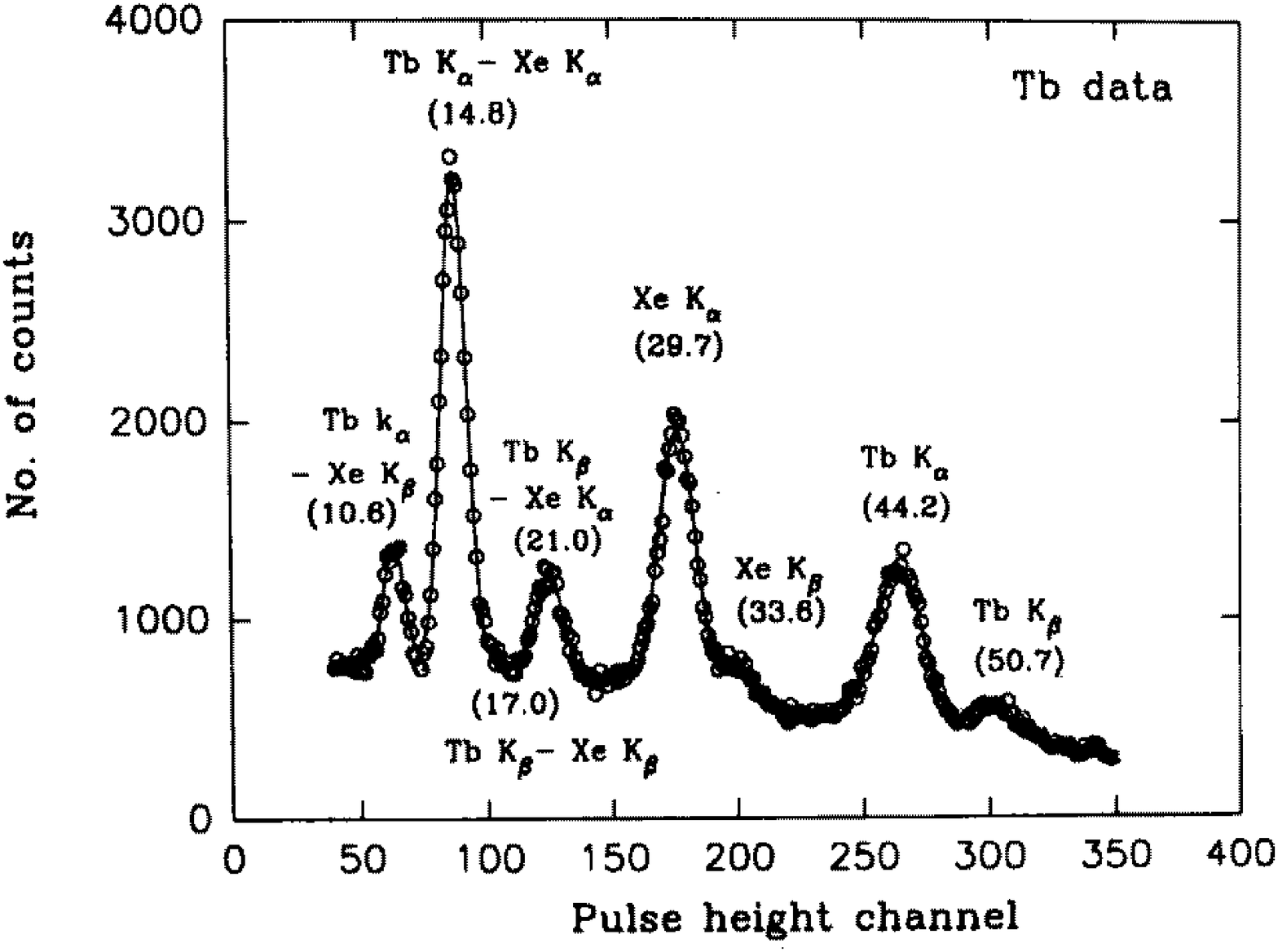}
Š\caption[]
{ 
 Response of the detector for characteristic X-rays from Tb radioactive
source. 
Various peaks including photo peaks and escape peaks are marked.
}
\end{figure}

\begin{figure}
\vskip 7cm
\includegraphics{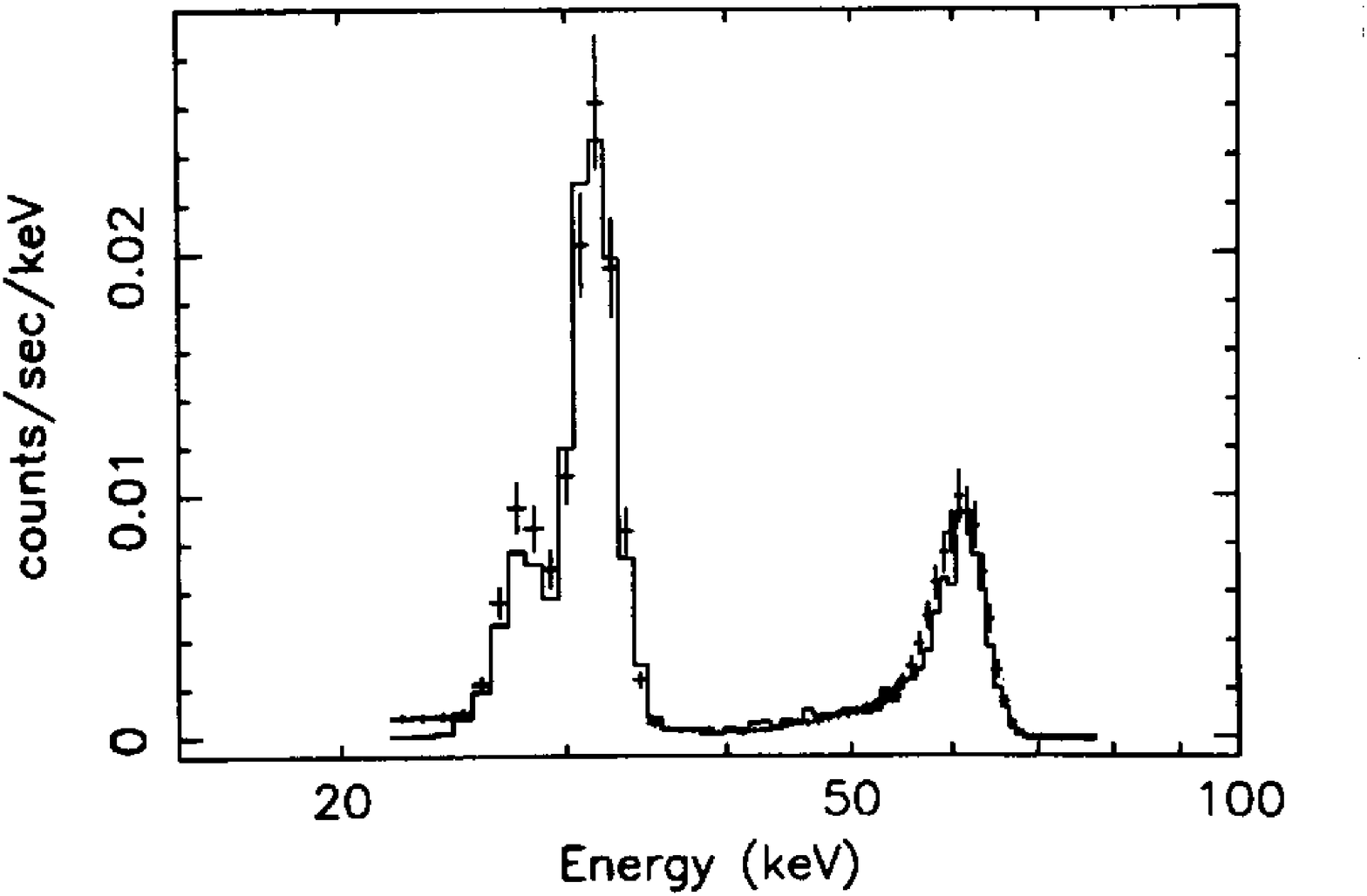}
Š\caption[]
{ 
 Response of layer 2 of detector A for Am$^{241}$ source. Data
points with error bars correspond to the observed pulse height distribution
and histogram corresponds to the predicted pulse height distribution for a
Gaussian line at 60 keV, convolved through the Monte Carlo simulated response
of the detector.
}
\end{figure}

\begin{figure}
\vskip 7cm
\includegraphics{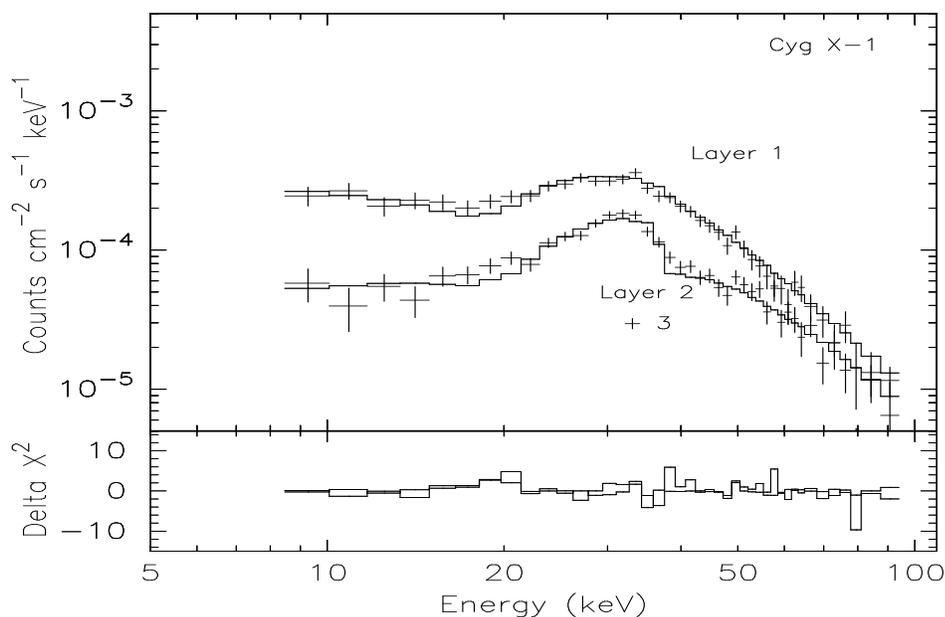}
Š\caption[]
{ 
Observed count rate spectra from Cyg X-1
 obtained from XMPC, shown separately
for layer 1 and bottom 2 layers. The best fit power-law model with photon
index ($\Gamma$) of 1.62, convolved through the detector response is shown
as histograms. The residuals to the model fit are shown in the lower panel 
of the figure as contribution to the $\chi^2$.
}
\end{figure}

\begin{figure}
\vskip 7cm
\includegraphics{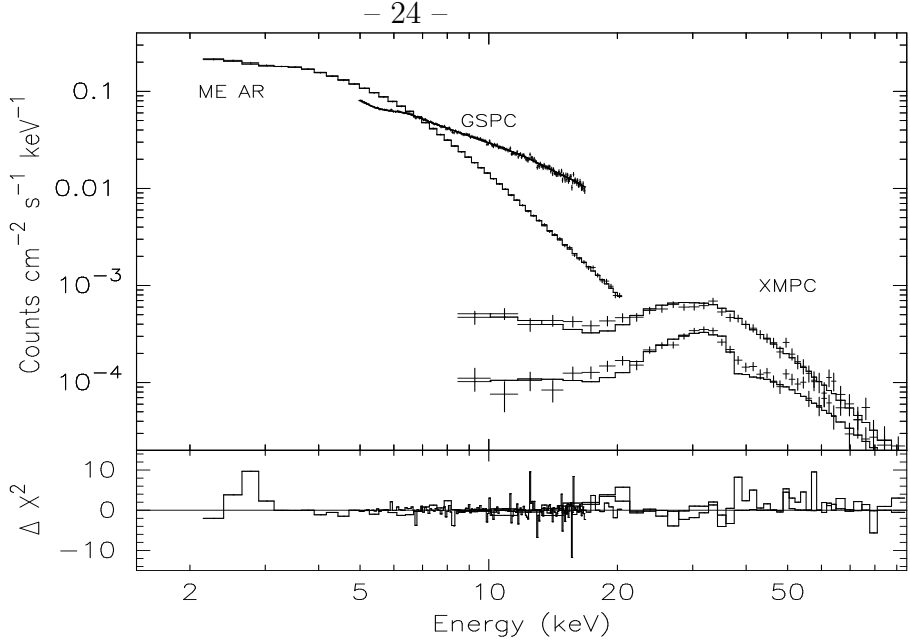}
Š\caption[]
{ 
 The observed count rate spectra from Cyg X-1 obtained from
 EXOSAT ME Ar and 
GSPC are shown along with the XMPC data. The best fit ionized reflection
model (along with low energy absorption, low energy blackbody, and a 
Gaussian line), obtained from a simultaneous fit to the three detectors,
is shown as histograms, after convolving through individual detector
response functions. The residuals are shown in the lower panel of the 
figure as contribution to the $\chi^2$.
}
\end{figure}

\begin{figure}
\vskip 7cm
\includegraphics{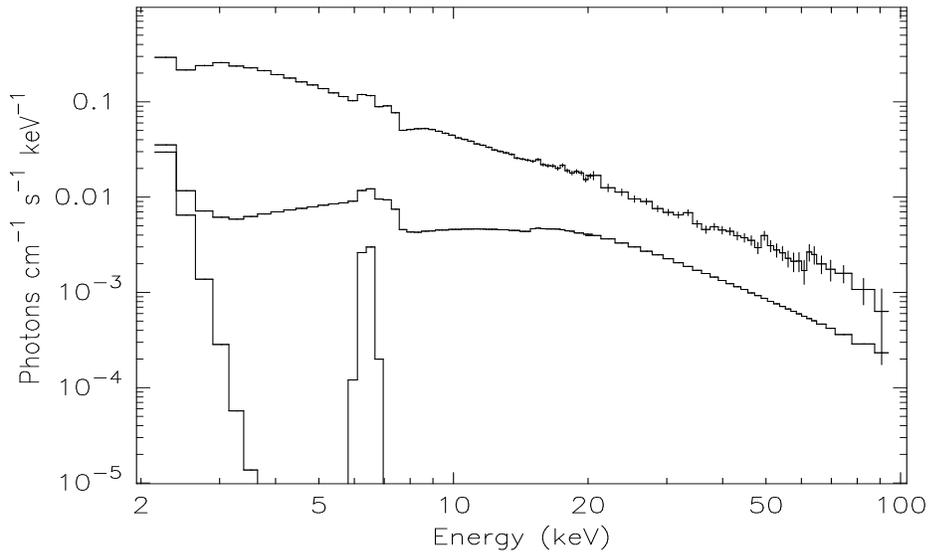}
Š\caption[]
{ 
 Deconvolved spectrum of Cyg X-1 obtained from Ar, GSPC and XMPC
detectors. For clarity, at a given energy, data from only one detector is
shown. The contribution from individual model components (blackbody, Gaussian
line and the reflection component) are shown as histograms.
}
\end{figure}

\begin{figure}
\vskip 7cm
\includegraphics{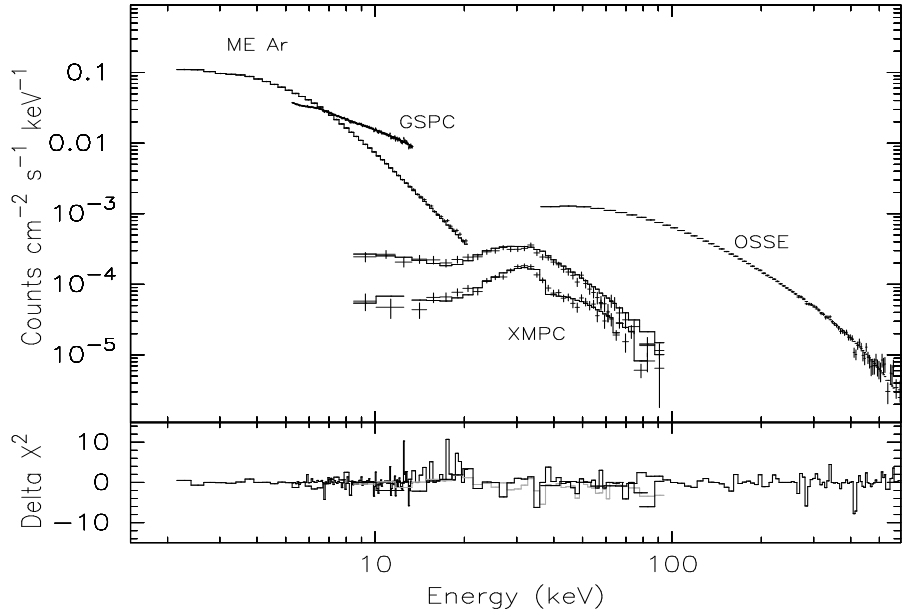}
Š\caption[]
{ 
Fig. 7 The observed count rate spectra from Cyg X-1 obtained from EXOSAT ME Ar, EXOSAT
GSPC, XMPC and OSSE are shown. The best fit model consisting of low energy
absorption, low energy blackbody, Gaussian line, absorption edge and two 
CompSTs convolved
through individual detector responses is shown as histogram. The residuals 
to the model fit are shown in the lower panel of the figure as contribution 
to the $\chi^2$.
}
\end{figure}

\begin{figure}
\vskip 7cm
\includegraphics{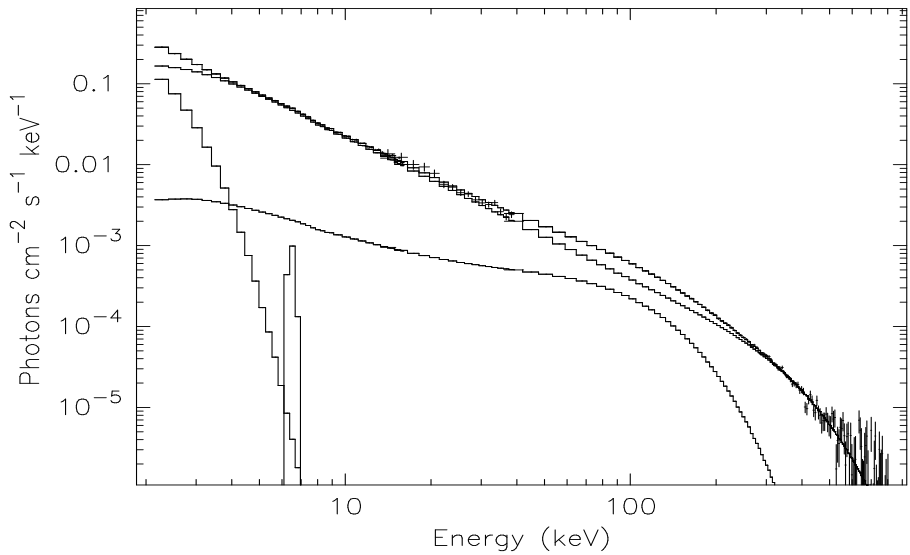}
Š\caption[]
{ 
The deconvolved spectra of Cyg X-1 obtained from EXOSAT ME Ar, XMPC
and OSSE are shown.
Contributions from individual model components (low energy blackbody of
temperature 0.328 keV; Gaussian line at 6.47 keV and two CompSTs of 
temperatures 80.2 keV and 29.7 keV, respectively) are shown separately
as histograms. 
}
\end{figure}
 
\end{document}